\documentstyle[prl,aps,epsf]{revtex}
\twocolumn
\tighten
\begin{document}
\draft
\twocolumn[
\hsize\textwidth\columnwidth\hsize\csname@twocolumnfalse\endcsname
\title{
Novel Quantum Spin Assisted Tunneling in Half Metallic
Manganite Tunnel Junctions
} 
\author{R. Y. Gu, L. Sheng, and C. S. Ting}
\address{
Texas Center for Superconductivity and Department of Physics, 
University of Houston, Houston, Texas 77204\\
}
\date{February 5, 2001}
\maketitle

\begin{abstract}
The electron tunneling in half metallic manganite tunnel
junctions is studied by using a quantum mechanically treated double
exchange model.
We show that the stimulation of spin excitations, caused by the 
strong Hund's coupling between the conduction $e_g$ electrons and the
localized 
quantum spins of the manganite ions, would assist electrons to tunnel
through the junction even with the antiparallel aligned magnetizations
of the electrodes. 
This mechanism gives rise to an extra tunnel
conductance, in addition to that predicted by Julliere's model. 
Our theory is in good agreement with the voltage dependence of the
tunnel conductance in manganite tunnel junctions observed by 
experiments.
\end{abstract} 
]

\narrowtext

Tunnel magnetoresistance (TMR) in ferromagnetic junctions (FJ), first
identified more than two decades ago \cite{julliere}, is of both
fundermental interest and
potential applications to magnetic sensors and memory devices. A simple
model \cite{julliere} proposed by Julliere is widely adopted to account
for the observed
TMR, in which the conductance of the polarized
electrons are proportional to the product of the density of states (DOS)
at the Fermi level from both magnetic electrodes for each spin
channel. According to this model, the TMR ratio is given by
\begin{equation}
\frac{\Delta R}{R}=\frac{R_{AP}-R_P}{R_P}=\frac{2P^2}{1-P^2}\ ,
\label{jul}
\end{equation}
where $R_{P}$ and $R_{AP}$ are the tunnel resistances 
when the magnetizations of the two ferromagnetic electrodes 
are in parallel (P) and antiparallel (AP) alignments. 
$P$ here is the spin polarization of the electrodes.
This model works quite well for Fe, Co and Ne based ferromagnets. From
Eq. (\ref{jul}) the TMR ratio tends to become infinity when the
electrodes are made of half-metallic ferromagnets where only one
spin component exists and $P=1$, 
such as in some mixed valence manganites La$_{1-x}$A$_{x}$MnO$_3$, (A=Ca,
Ba, Sr, etc.). The bulk of these manganese oxides, well known as the
colossal magnetoresistance (CMR) materials, have been intensively studied
in recent years. The involved physics is 
based on the double exchange
(DE) model \cite{zener,varma,furukawa} which can be described by the 
following Hamiltonian
\begin{equation}
H=-t\sum_{ij, \sigma} c^{\dag}_{i\sigma}c_{j\sigma}-J_H\sum_i 
\vec{S}_i\cdot \vec{\sigma}_i \ ,
\label{de}
\end{equation}
where the two terms represent the conduction $e_g$ electron hopping
between neighboring Mn ions and the 
on-site Hund's coupling between the electron spin $\vec{\sigma}$ and the
localized
$t_{2g}$ core spin $\vec{S}_i$ of Mn$^{3+}$. 
It is well known that in mixed valence manganite system the strong Hund's
coupling $J_H>>t$ dominates the basic physics 
by separating the two spin subbands. The even stronger on-site
Coulomb repulsion (not written out in Eq. (\ref{de})) further enhances the
energy difference between the two
subbands. As a result,
a reasonable assumption of a full spin polarization
($P=1$) was extensively accepted in the literatures. For this reason,
the manganese oxides have been recognized as being good candidates for
the study of various spin polarized tunnelings
\cite{sun,lu,viret,obata,jo}.  
The authors in Refs.\cite{lu,viret,obata}
fabricated FJ using
La$_{0.67}$Sr$_{0.33}$MnO$_3$, La$_{0.8}$Sr$_{0.2}$MnO$_3$ and
La$_{0.7}$Sr$_{0.3}$MnO$_3$ to obtain a high TMR ratio 80\%, 150\%, and
430\%
respectively \cite{lu,viret,obata}.
More recently, a TMR
ratio as high as 730\% are obtained by Jo et al. with
La$_{0.7}$Ca$_{0.3}$MnO$_{3}$ as electrodes \cite{jo}. All these
results were 
explained by Julliere's model, from which a spin polarization $P=0.54,
0.65, 0.83$ and $0.88$ are respectively assumed. While these values
are much larger than that of a typical common magnetic transition metal,
they are
still far away from the expected $P=1$ for half-metallic
manganites. Furthermore, it is found that the tunnel conductance $G$ 
in these manganite junctions depends on the bias voltage $V$. 
In both the P and AP
magnetization alignments, $G$ has a considerable increase when $V$ rises
up to tens of mV \cite{sun,lu,viret,obata,jo}. As a result, TMR is
strongly reduced.
Bratkovsky attributed the $V$-dependence of $G$ to the lowering of the 
barrier height of the insulator layer by the applied voltage, but his
calculation shows that this
effect becomes appreciable only when $|eV|>0.3e\makebox{V}$ 
\cite{bratkovsky}.
Therefore the origin of such a voltage dependence of the conductance can
not be
explained by the existing theories \cite{julliere,bratkovsky} and so far
still is a puzzle.
To account for the discrepancy between the expected full spin polarization
of the observed finite TMR ratio, 
it was proposed that 
the presence of defect states in the
barrier or resonant state can reduce the TMR \cite{bratkovsky}. 
A reduced TMR was also
obtained by Lyu et al. who  introduced spin-flip matrix elements in the 
tunneling Hamiltonian \cite{lyu}. 
Recently, Itoh et al. also studied the effect of the temperature
dependence of the spin polarization on TMR \cite{itoh}.

In this work, we investigate the voltage-dependent electronic
tunneling in the
half-metallic manganite junctions
based on a quantum mechanical treatment of the DE model.
It is found
that in the AP aligned electrodes even 
without any spin-flip scattering from the tunneling Hamiltonian, an
electron can still tunnel through the junction resulting in a
nonzero conductance and thus finite TMR. Such a tunneling
comes from 
the intrinsic quantum spin effect, and is accompanied by the
stimulation of spin excitations in the ferromagnetic electrodes. 
In the classical spin $S\rightarrow \infty$ limit, our result reduces to a
formula similar to that of Julliere's model.
For finite $S$, 
the quantum spin effect becomes prominent and suppresses strongly
the TMR amplitude.
We show that our theory explains well the voltage dependence of the
tunnel conductance in manganite tunnel junctions observed in
Refs.{\cite{lu}} and \cite{obata}.

We start from the following model for manganite tunnel junctions
\begin{equation}
H=H_L+H_R+H_T\ ,
\end{equation}
where $H_L$ and $H_R$ are the quantum DE Hamiltonian with the form of
Eq. (\ref{de}) in the left and right electrodes. $H_T$ 
represents the spin-conserved incoherent tunneling process of 
conduction electrons
through the insulator barrier. In the momentum space, it is given by
\begin{equation}
H_T=\sum_{pp^{\prime},\sigma}(T_{pp^{\prime}}c^{R\dag}_{p^{\prime}\sigma}c^L_{p\sigma}+h.c.)
\ .
\end{equation}

According to the perturbation theory, the net electronic tunnel current
is 
\begin{eqnarray}
I=\frac{2\pi e}{\hbar}{\cal Z}^{-1}\sum_{m,n}\sum_{pp^{\prime},\sigma}
|\langle
n|T_{pp^{\prime}}c^{R\dag}_{p^{\prime}\sigma}c^L_{p\sigma}|m\rangle|^2
\nonumber \\
\times(\makebox{e}^{-\beta E_m}-\makebox{e}^{-\beta E_n})
\delta(E_n-E_m+eV)\ ,
\label{tunneling}
\end{eqnarray}
where $|m\rangle$, $|n\rangle$ are eigenstates of $H_L+H_R$ with
energy $E_m$ and $E_n$, ${\cal Z}=$Tr$\makebox{e}^{-\beta(H_L+H_R)}$, with 
$\beta=(k_BT)^{-1}$ the inverse temperature, and $V$ is the applied bias
voltage. 
The delta function
ensures the conservation of energy in the tunneling process. 

To derive the tunnel current from Eq. (\ref{tunneling}),
it is convenient to 
use the Schwinger-boson or slave-fermion representation 
\cite{auerbach,sheng}
in which the electron operator $c_{i\sigma}$ is expressed as a combination
of a spinless charge fermion $f_i$ and a neutral spin
boson $b_{i\sigma}$. 
In the present strong Hund's coupling ($J_H\rightarrow 
\infty$) case,
the spin of an electron on a site is aligned parallelly
to the local spin, and the electron
operator can be written as
$c^L_{i\sigma}=f^L_ib^L_{i\sigma}/\sqrt{2S_l+1}$ for $i\in L$, and
$c^R_{i\sigma}=f^R_i\sum_{\sigma^{\prime}}U_{\sigma\sigma^{\prime}}(\theta)
b^R_{i\sigma^{\prime}}/\sqrt{2S_l+1}$
for $i\in R$, where $S_l=3/2$ is the local spin at the site $i$,  
and $U_{\sigma\sigma^{\prime}}(\theta)$ are matrix elements of the
$2\times 2$
matrix $U(\theta)=\cos(\theta/2)\hat{{\bf 1}}+i\sin(\theta/2)\sigma_y$. 
Without loss of
generality, we assume the magnetization in the left electrode is
parallel to the z-axis while the magnetization in the right electrode
makes an arbitrary
angle $\theta$ with the z-axis. $\theta=0$ and $\pi$ correspond to
the P and AP configurations of the junction magnetizations.
The total spin of a site (the local spin plus the spin of the
itinerant electron) is
$\vec{S}^{tot}_i=\frac{1}{2}
\sum_{\sigma\sigma^{\prime}}b^{\dag}_{i\sigma}
\vec{\sigma}_{\sigma\sigma^{\prime}}b_{i\sigma^{\prime}}$, with the
local constraint
$\sum_{\sigma}b^{\dag}_{i\sigma}b_{i\sigma}=2S_l+f_i^{\dag}f_i$.  
We use a mean-field approximation that was derived in
Ref.\cite{sheng} for the
unperturbed Hamiltonian $H_{\alpha}$ ($\alpha=L, R$) 
\begin{equation}
H_{\alpha}=\sum_{k}\epsilon_{k}f^{\alpha\dag}_kf^{\alpha}_k
+\sum_k\omega_kb^{\alpha\dag}_{k\sigma}b^{\alpha}_{k\sigma}\ ,
\label{h0}
\end{equation}
where $\omega_k=\rho k^2$, with the spin stiffness
$\rho=\rho_0\{1-\sum_k\gamma_k[\makebox{exp}
(\beta\rho_0\gamma_k)-1]^{-1}/12S\}$, and
$\epsilon_{k}=\bar{t}(\gamma_k-6)$ with $\bar{t}=t\sum_{\sigma}\langle
b^{\dag}_{i\sigma}b_{j\sigma}\rangle/2S$.
Here $\rho_0=t\langle f^{\dag}_if_j\rangle/2S$, 
$\gamma_k=6-2(\cos k_x+\cos k_y+\cos k_z)$ and $S=S_l+(1-x)/2$ is
the average spin per site. 
The total number of spin bosons is $n_{\uparrow}+n_{\downarrow}=2NS$,
where $n_{\sigma}=\sum_{k}\langle b_{k\sigma}^{\dag}b_{k\sigma}\rangle$
and $N$ is
the total number of sites. 
At temperatures below the Curie temperature
$T_c$ and at each electrode, there are macroscopic numbers of bosons
condense into $\vec{k}=0$ and
$\sigma=\uparrow$ state,
$n_{0,\uparrow}=n_{\uparrow}-n_{\downarrow}=2NSm$, where
$m=(n_{\uparrow}-n_{\downarrow})/(n_{\uparrow}+n_{\downarrow})=M/M_s$,
with $M$ the magnetization of the electrode and $M_s$ its
saturation value, is just the normalized magnetization of the electrode.
Employing the Schwinger boson representation
for Eq. (\ref{tunneling}) and after some algebra, we obtain the
following tunnel current
\begin{eqnarray}
&I&=\frac{2\pi e|\bar{T}|^2}{\hbar(2S_l+1)^2N^2}\sum_{qq^{\prime}}
\sum_{k\sigma,k^{\prime}\sigma^{\prime}}|U_{\sigma\sigma^{\prime}}(\theta)|^2
\nonumber \\
&\times&[n^f_qn^b_{k\sigma}(1-n^f_{q^{\prime}})(n^b_{k^{\prime}\sigma^{\prime}}
-(1-n^f_q)(n^b_{k\sigma}+1)n^f_{q^{\prime}}n^b_{k^{\prime}\sigma^{\prime}}]
\nonumber \\
&\ &\times
\delta(\epsilon_q+\omega_k-\epsilon_{q^{\prime}}-\omega_{k^{\prime}}+eV)
\ ,
\label{current} 
\end{eqnarray}
where $n^f_q=[\makebox{e}^{\beta(\epsilon_q-\epsilon_F)}+1]^{-1}$ and
$n^b_{k\sigma}=(\makebox{e}^{\beta\omega_k}-1)^{-1}$ are the distribution
functions of charge fermions and spin bosons.
In the derivation we have treated 
$|T_{pp^{\prime}}|^2$ as a constant $|\bar{T}|^2$.
Taking into account the condensation 
of the spin bosons in the $\vec{k}=0$ and 
$\sigma=\uparrow$ state,  and that the band width of the
fermion is in the order of electron volts much larger than energy
scales of $k_BT$ and $eV$, thus the DOS of fermions near the Fermi
level is approximated as a constant, the
dynamical conductance of the junction $G=dI/dV$ can be derived as
\begin{eqnarray}
&G&(\theta)=\frac{1}{2}G_0\{
1+m^2\cos\theta+\frac{1-m}{S}
+\frac{m}{S}\int_0^{e|V|} g_b(\omega)d\omega 
\nonumber
\\
&+&\frac{1}{S^2}\int_0^{\omega_m}g_b(\omega)d\omega  
\int_0^{\omega+e|V|}
[n^b(\omega)-n^b(\omega^{\prime})]
g_b(\omega^{\prime})d\omega^{\prime} \} \ ,
\label{conductance}
\end{eqnarray}
where $G_0=\frac{2\pi
e^2}{\hbar}(\frac{2S}{2S_l+1})^2|\bar{T}|^2g_f^2(\epsilon_F)$ 
with $g_f(\epsilon)$ the DOS of fermions,
$n^b(\omega)=(\makebox{e}^{\beta\omega}-1)^{-1}$,
$\omega_m=(6\pi^2)^{2/3}\rho$ is the maximum energy of the spin
excitation, and 
$g_b(\omega)$ is the normalized DOS of bosons, i.e.,
$\int_0^{\omega_m}g_b(\omega)d\omega=1$. 

Eq. (\ref{conductance}) is the tunnel conductance of the half-metallic
manganite for
arbitrary magnetization configuration, in which
a voltage dependence is found in the last two 
terms. All the last three terms are related to the spin size $S$ and
vanish when the spin is treated classically ($S\rightarrow \infty$).
This indicates that they come from the quantum spin effect.
In the classical spin limit 
the TMR reduces to $\Delta R/R=2m^2/(1-m^2)$, which has
a similar form to Julliere's formula Eq. (\ref{jul}), but the
spin polarization $P$ there is replaced by
the normalized magnetization $m$ here. 
$P$ and $m$ are usually different, 
the former is the relative difference of the DOS at the Fermi
level between the two electron subbands split by the Hund's coupling,
in the large $J_H$ limit it is always 1 (see, e.g., the calculation 
of $P$ in Ref. \cite{itoh}) so the system is half-metallic, 
while the latter reaches 1 only at zero temperature, 
and decreases with the increase temperature due to the spin wave
excitations. When the quantum spin effect
is taken into account, even at zero temperature with full magnetization
(m=1), the fourth term contributes a nonzero conductance.
In the application of a small voltage $e|V|<\omega_m$, this term reduces
to $(e|V|/\rho)^{3/2}/6\pi^2S$, so that our theory predicts the
tunnel conductance $G(\theta)$ at low temperatures behaves as
$A+B|V|^{3/2}$,  where $A, B$ are constants independent of $V$. 
In Fig. 1(a) we show a comparison between our theory and experimental
data
in Ref.\cite{lu} in the AP
magnetization configuration of the
La$_{0.67}$Sr$_{0.33}$MnO$_3$/SrTiO$_3$/La$_{0.67}$Sr$_{0.33}$MnO$_3$ 
junction at $T=4.2$K. The electronic hopping $t$ is taken to be 
0.35 eV so that $\rho\approx 14$meV.
Apart from the difference of a small voltage-independent
constant, which 
may come from spin-flip centers, defect states in the barrier, $m<1$ at
finite temperature or imperfect alignment of moments in a less-than-ideal 
sample\cite{bratkovsky,lyu}, our calculated
conductance is in excellent agreement with the 
experimental measurement. For comparison the conductance in the
classical spin treatment ($S=\infty$) of the DE model is also plotted 
by the dashed line, which is independent of V. 
The dependence of the conductance on voltage also indicates that we have
to specify the value of the bias voltage when discussing 
the TMR. Interestingly, for the case of $e|V|$ larger than the
band width of spin excitations, both the fourth and the last terms in
Eq. (\ref{conductance}) become independent of the bias
voltage. While further increasing of the conductance with the rise
of the voltage may be still possible through lowering the energy barrier
of the insulator layer \cite{bratkovsky}, 
such an increase does not
change the ratio of TMR. From Eq. (\ref{conductance}) and under 
this large voltage,
the TMR ratio in the full magnetization $m=1$ of an ideal 
half metallic manganite junction is $\Delta R/R=2S$, which is less than
$400\%$ for realistic $S<2$.  
A divergent TMR ratio can be obtained only when $S$ is treated as
classically ($S\rightarrow \infty$) so that the quantum spin effect
induced
tunneling vanishes and
the conductance in the AP magnetization configuration goes to zero. 

The effect of temperature comes into the conductance mainly through the
normalized magnetization $m$, which decreases with increasing the 
temperature. In Fig. 1(b) we plot the voltage-dependent conductance at
different temperatures. To compare our calculations with the
experimental results in Ref.\cite{obata} 
in which the FJ was 
La$_{0.8}$Sr$_{0.2}$MnO$_3$/SrTiO$_3$/La$_{0.8}$Sr$_{0.2}$MnO$_3$, 
we need to determine the magnetizations at different temperatures. Here
the magnetizations are estimated from the measurement in Ref.\cite{park}, 
they are taken as $m=1$ at $T=0$, $m=0.6$ at $T=0.4T_c$, and $m=0.2$
at $T=0.8T_c$, with $T_c=290$K\cite{obata}. 
In Fig. 1(b) at each temperature we
shift the voltage-independent constant conductance so that
the comparison of the 
$V$-dependent conductance at different temperatures can be shown more 
clearly.
Fig. 1(b) shows that the increase of the conductance with the increasing 
voltage becomes slower at higher temperatures,
in agree with the experimental results of Ref.\cite{obata}. 

\begin{figure}[h]
\hspace{-2cm}
\centerline{epsfxsize=3.0truein \epsfbox{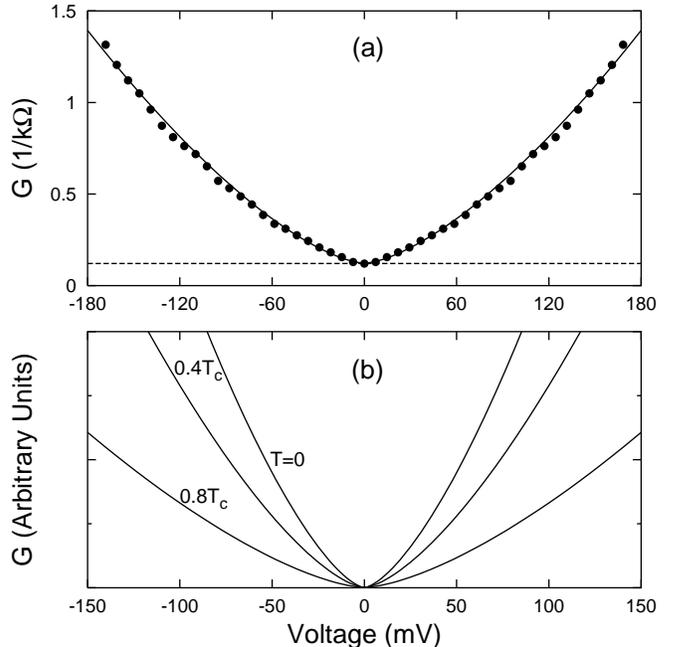}}
\caption[*]{
Tunnel conductance $G(\pi)$ 
as a function of the applied voltage 
(a) at temperature $T=4.2$K and doping $x=0.33$, 
where the conductance has been shifted a small voltage-independent
constant so as to compare with the experimental measurement of
Ref.\cite{lu}, which is shown by the dots in the present figure, and
$G_0$
in Eq. (\ref{conductance}) is taken to be 6.11 $(k\Omega)^{-1}$, the
dashed
line is for $S=\infty$, which is voltage independent,
(b) at temperatures $T=0, 0.4T_c$ and $0.8T_c$ and $x=0.2$, with
$T_c=290$K.
The electron hopping is taken to be $t=0.35$eV. 
}
\end{figure}

At this stage, we wish to address on  why it 
is possible for electrons to tunnel through the junction 
for a spin conserved tunneling Hamiltonian
when the
magnetizations of the two fully polarized electrodes are antiparallel to 
each other. Although the Schwinger boson method is convenient for
calculation, the spin boson operator in it describes the total spin
on a site as a whole, which conceals the actual motion of different spins. 
To see the physical origin of the nonvanishing $G(\pi)$, we now 
start directly from the DE Hamiltonian Eq. (\ref{de}). 
At each site the Hund's coupling 
$-J_H\vec{S}_i\cdot\vec{\sigma_i}$ has two eigenvalues
$-J_HS_l$ and $J_H(S_l+1)$, where the former is 
$2S_l+2$ fold degeneracy with total spin 
$S_l+1/2$, and the latter is $2S_l$ degeneracy with total spin $S_l-1/2$.
In the case of the strong Hund's coupling, at an occupied site only 
the lower energy state can exist, i.e., 
the spin of the electron must be parallel to the local spin.
In this case, 
the effective DE hopping of the $e_g$ electrons between two
neighboring sites is usually treated to be $t_{eff}=t\cos\theta_{ij}$
\cite{zener}, where
$\theta_{ij}$ is the relative angle between local spins on the two sites. 
When the effective hopping is generalized to 
manganite junctions with AP configuration at zero temperature, 
$\theta_{ij}=\pi$ for $i$ and $j$ in different electrodes, 
and the electron can not tunnel through the junction.
That is exactly what Julliere's model predicts.
However, such a classical picture neglects the quantum nature
of local spins. 
In fact, when the quantum effect is taken into
account, the spin of a conduction electron in one electrode can not be
perfectly antiparallel to a localized spin in the other electrode. 
For example, when a spin-up
($s_z=1/2$) electron in the left electrode moves to a spin-down site
($S_z=-S_l$) in the right electrode, the formed 
$|1/2,-S_l\rangle$ state, which is not an eigenstate of the Hund's
coupling,
has nonzero components in both the spin parallel 
(the $e_g$ electron spin and the local spin are in parallel)
and spin antiparallel subspaces. Using the method of projection 
operator, it is straightforward to show 
\begin{equation}
|\frac{1}{2},-S_l\rangle=\sqrt{\frac{1}{2S_l+1}}|\psi_p\rangle+
\sqrt{\frac{2S_l}{2S_l+1}}|\psi_{ap}\rangle\ ,
\label{decompo}
\end{equation}
where $|\psi_p\rangle=
(|1/2,-S_l\rangle+\sqrt{2S_l}|-1/2,-S_l+1\rangle)/\sqrt{2S_l+1}$
and $|\psi_{ap}\rangle=
(\sqrt{2S_l}|1/2,-S_l\rangle-|-1/2,-S_l+1\rangle)/\sqrt{2S_l+1}$
are states in which the electron spin and the local spin are in parallel
and antiparallel.
Both $|\psi_p\rangle$ and $|\psi_{ap}\rangle$ are eigenstates of 
$\hat{S}_{tot}^{z}$ (z-component of the total spin of the site)
with eigenvalues $-S_l+1/2$. At the same time, they  
are also eigenstates of the Hund's coupling 
$-J_H\vec{S}_i\cdot\vec{\sigma_i}$, with eigenvalues  
$-J_HS_l$ for $|\psi_p\rangle$ and $J_H(S_l+1)$ for $|\psi_{ap}\rangle$. 
For large $J_H$, state $|\psi_{ap}\rangle$ is forbidden due to its
very high energy. However, 
Eq. (\ref{decompo}) shows that $|1/2,-S_l\rangle$ has a nonzero component
in $|\psi_p\rangle$, i.e.,  
there is a  finite probability $1/(2S_l+1)$ 
for the spin of a spin-up electron to be 
parallel to a spin-down local site, so that even in this 
$J_H\rightarrow\infty$ case the electron may still 
tunnel through the junction. This tunneling process
vanishes when $S_l\rightarrow\infty$, which is an implication 
that this effect is induced by the quantum nature of the localized spins.
In the second term of $|\psi_{p}\rangle$, the spin of the
electron is flipped while the local spin of the site
is twisted. Such a spin-flip and spin-twist process is an intrinsic
dynamic effect originated from the Hund's coupling
and the discrete quantum spin orientations.
Moreover, the twist of the local spin
can spread to the whole system to further lower the energy,
thus stimulates a spin excitation and
enables electrons to tunnel through the barrier. 
This mechanism is very different from that considered in some other 
papers \cite{lyu,zhang,gu}.  
The energy scale 
involved in this tunneling process is the spin excitation energy 
which comes from the applied voltage and thus causes the induced
tunneling to be voltage dependent.

Finally, it is worth to mention that in some experiments, at temperatures
77K and above, a quadratic variation of the conductance with the applied
voltage were found \cite{viret,jo}. This is likely due to
the fact that the DOS of the spin excitations near the
ferromagnet/insulator
(F/I) regions is different from that in the bulk manganites. Experimental
study of the 
magnetic properties at surface boundary of the half-metallic ferromagnet
La$_{0.7}$Sr$_{0.3}$MnO$_3$ indeed indicates that except at very low
temperature ($T<30$K), the temperature dependence of the surface-boundary
magnetism is
significantly different from that of the bulk \cite{park}.
In the present theory, a $V^2$ dependence of the
conductance can be produced if the spin excitation spectrum has the form 
$\omega_k=\rho_{\|}k_{\|}^2+\rho_{\perp}|k_{\perp}|$,
where $k_{\|}$ and $k_{\perp}$ are the wave vectors in and
perpendicular to the plane of the F/I interface, respectively. 
Such a
dispersion relation was proposed for the lightly doped 
La$_{1-x}$Sr$_{x}$MnO$_3$ with layered antiferromagnetic spin alignment
\cite{woodfield}, and it may exist in the F/I interface
regions of the half metallic manganite junctions.


{\bf Acknowlegement} - This work is supported by a grant from Texas ARP
grant (ARP-003652-0241-1999), the Robert A. Welch Foundation and the Texas
Center for Superconductivity at the University of Houston.


\begin{references}

\bibitem{julliere} M. Julliere, Phys. Lett.{\bf 54A}, 225 (1975).

\bibitem{zener} C. Zener, Phys. Rev. {\bf 82}, 403 (1951); P. W. Anderson
and H. Hasegawa, {\it ibid} {\bf 100}, 675 (1955); P. -G. de Gennes, {\it
ibid} {\bf 118}, 141 (1960).

\bibitem{varma} C. M. Varma, Phys. Rev. B {\bf 54}, 7328 (1996); E. 
M\"{u}ller-Hartmann and E. Daggoto, {\it ibid}, R6814 (1996).

\bibitem{furukawa} N. Furukawa, J. Phys. Soc. Jpn. {\bf 63}, 3214
(1994); J. Inoue and S. Maekawa, Phys. Rev. Lett. {\bf 74}, 3407 (1995);
A. J. Millis, P. B. Littlewood and B. I. Shraiman, Phys. Rev. Lett. {\bf
74}, 5144 (1995).

\bibitem{sun} J. Z. Sun, W. J. Gallagher, P. R. Duncombe,
L. Krusin-Dlbaum, R. A. Altman, A. Gupta, Yu Lu, G. Q. Gong and Gang
Xiao,
Appl. Phys. Lett. {\bf 69}, 3266 (1996).

\bibitem{lu} Y. Lu, X. W. Li, G. Q. Gong, G. Xiao, A. Gupta, P. Lecoeur,
J. Z. Sun, Y. Y. Wang, and V. P. Dravid, Phys. Rev. B{\bf 54}, R8357
(1996).

\bibitem{viret} M. Viret, M. Drouet, J. Nassar, J. P. Contour, C. Fermon
and A. Fert, Europhys. Lett. {\bf 39}, 545 (1997).

\bibitem{obata} T. Obata, T. Manako, Y. Shimakawa, and Y. Kubo,
Appl. Phys. Lett. {\bf 74}, 290 (1999). 

\bibitem{jo} M. Jo, N. D. Mathur, N. K. Todd, and M. G. Blamire,
Phys. Rev. B{\bf 61}, 14905 (2000).

\bibitem{bratkovsky} A. M. Bratkovsky, Phys. Rev. B{\bf 56}, 2344 (1997);
A. M. Bratkovsky, Appl. Phys. Lett. {\bf 72}, 2334 (1998).

\bibitem{lyu} P. Lyu, D. Y. Xing and J. Dong, Phys. Rev. B{\bf 58},
54 (1998).

\bibitem{itoh} H. Itoh, T. Ohsawa and J. Inoue, Phys. Rev. Lett.{\bf 84},
2501 (2000).

\bibitem{auerbach} See A. Auerbach, {\it Interacting Electrons and Quantum
Magnetism}, (Springer-Verlag, Berlin, 1994).

\bibitem{sheng} L. Sheng, H. Y. Teng, and C. S. Ting, Phys. Rev. B{\bf
58}, 8186 (1998).

\bibitem{park} J.-H. Park, E. Vescovo, H. -J. Kim, C. Kwon, R. Ramesh, and
T. Venkatesan, Phys. Rev. lett.{\bf 81}, 1953 (1998).

\bibitem{zhang} S. Zhang, P. M. Levy, A. C. Marley and S. S. P. Parkin,
Phys. Rev. Lett.{\bf 79}, 3744 (1997).

\bibitem{gu} R. Y. Gu, D. Y. Xing, and J. Dong, J. Appl. Phys. {\bf 80},
7163 (1996).

\bibitem{woodfield} B. F. Woodfield, M. L. Wilson, and J. M. Byers, Phys.
Rev. Lett. {\bf 78}, 3201 (1997).

\end{references}
\end{document}